\documentclass[%
 reprint,
nobibnotes,
 amsmath,amssymb,
 aps,
]{revtex4-1}

\usepackage{graphicx}
\usepackage{dcolumn}
\usepackage{bm}
\usepackage{braket}
\usepackage{amsmath}
\usepackage{amssymb}
\usepackage{enumerate}
\usepackage{booktabs}
\usepackage{subfigure}
\usepackage[update,prepend]{epstopdf} 
\usepackage{longtable}

\usepackage[english]{babel}
\usepackage{color}
\usepackage{listings}

\makeatletter
\def\hlinewd#1{%
	\noalign{\ifnum0=`}\fi\hrule \@height #1 %
	\futurelet\reserved@a\@xhline}
\makeatother

\begin{document}

\preprint{APS/123-QED}

\title{Optimizing Quantum Circuits for Arithmetic}

\author{Thomas H\"aner}
\email{haenert@phys.ethz.ch}
\affiliation{Microsoft Quantum, Microsoft, Redmond, WA 98052, USA}
\affiliation{Institute for Theoretical Physics, ETH Zurich, 8093 Zurich, Switzerland}

\author{Martin Roetteler}
\email{martinro@microsoft.com}
\affiliation{Microsoft Quantum, Microsoft, Redmond, WA 98052, USA}

\author{Krysta M.~Svore}
\email{ksvore@microsoft.com}
\affiliation{Microsoft Quantum, Microsoft, Redmond, WA 98052, USA}

\date{\today}

\renewcommand{\paragraph}[1]{\vspace{5pt}\noindent\textbf{#1}}
\graphicspath{{images/}}

\begin{abstract}
Many quantum algorithms make use of oracles which evaluate classical functions on a superposition of inputs. In order to facilitate implementation, testing, and resource estimation of such algorithms, we present quantum circuits for evaluating functions that are often encountered in the quantum algorithm literature. This includes Gaussians, hyperbolic tangent, sine/cosine, inverse square root, arcsine, and exponentials. We use insights from classical high-performance computing in order to optimize our circuits and implement a quantum software stack module which allows to automatically generate circuits for evaluating piecewise smooth functions in the computational basis. Our circuits enable more detailed cost analyses of various quantum algorithms, allowing to identify concrete applications of future quantum computing devices. Furthermore, our resource estimates may guide future research aiming to reduce the costs or even the need for arithmetic in the computational basis altogether.
\end{abstract}

\pacs{Valid PACS appear here}
                              
\maketitle


\section{\label{sec:intro}Introduction}

Quantum computers are expected to excel at certain computational tasks with an asymptotic advantage over their classical counterparts. Examples for such tasks include factoring~\cite{shor1994algorithms} and the simulation of quantum chemical processes~\cite{reiher2016elucidating,babbush2016exponentially}. While new quantum algorithms tackling these problems offer favorable asymptotic behavior, exact runtime estimates are often lacking due to the absence of reversible implementations for functions such as the ones considered in this paper. However, the implementation details of these functions greatly influence the constant overheads involved and, thus, also the crossover points at which the choice of quantum/classical algorithm changes.

We address this issue by presenting circuits for arithmetic which can be added to a quantum software stack such as LiQ$Ui\Ket{}$~\cite{wecker2014liqui}, Quipper~\cite{green2013quipper}, ScaffCC~\cite{javadiabhari2014scaffcc}, Q\#~\cite{svore2018q}, and ProjectQ~\cite{steiger2016projectq} to name a few. In particular, we discuss the implementation of general smooth functions via a piecewise polynomial approximation, followed by functions that are used in specific applications. Namely, we analyze the costs of implementing an inverse square root ($1/\sqrt x$) using a reversible fixed-point version of the method used in the computer game Quake III Arena~\cite{quake3} and we then combine this with our evaluation scheme for smooth functions in order to arrive at an implementation of $\arcsin(x)$.

Having reversible implementations of these functions available enables more detailed cost analyses of various quantum algorithms such as
HHL~\cite{harrow2009quantum}, where the inverse square root can be used to arrive at $x\mapsto 1/x$ and $\arcsin(x)$ can be used to get $1/x$ from the computational basis state into the amplitude. Similar use cases arise in Quantum Metropolis sampling~\cite{KOV+:2011}, Gibbs state preparation~\cite{PW:2009} and in the widely applicable framework of Quantum Rejection Sampling~\cite{ORR:2013} to transform one or more samples of a given quantum state into a quantum state with potentially different amplitudes, while maintaining relative phases. In all these examples the computation of $\arcsin(x)$ is useful for the rejection sampling step. Further applications of numerical functions can be anticipated in quantum machine learning, where sigmoid functions may need to be evaluated on a superposition of values employing $\tanh(x)$, and $1/\sqrt x$ can be used for (re-)normalization of intermediate results~\cite{CNW:2010}. In quantum algorithms for chemistry, further examples for numerical functions arise for on-the-fly computation of the one- and two-body integrals~\cite{babbush2016exponentially}. There, $1/\sqrt x$ as well as the evaluation of smooth functions such as Gaussians is needed. Similarly, on-the-fly computation of finite element matrix elements often involves the evaluation of functions such as $\sin(x)$ and $\cos(x)$~\cite{Scherer2017}.

\paragraph{Related work.} As a result of the large impact that the implementation details of such functions may have on the practicality of a given quantum algorithm, there is a vast amount of literature available which provides circuits for various low-level arithmetic functions such as addition~\cite{takahashi2009quantum,draper2000addition,cuccaro2004new,Draper:2006:LQC:2012086.2012090}. Furthermore, Refs.~\cite{cao2013,Bhaskar:2016:QAC:3179448.3179450,MT:2018} discuss implementations of higher-level arithmetic functions such as $\sin(x)$, $\arcsin(x)$ and $\sqrt{x}$ which we also consider in the present work, although using different approaches. In particular, our piecewise polynomial evaluation circuit enables evaluating piecewise smooth functions to high accuracy using polynomials of very low degree. As a result, we require only a small number of additions and multiplications, and few quantum registers to hold intermediate results in order to achieve reversibility. While Ref.~\cite{cao2013} employs several evaluations of the $\sin(x)$ function in order to hone in on the actual value of its inverse, our implementation of $\arcsin(x)$ features costs that are similar to just one invocation of $\sin(x)$ for $x\in [-0.5,0.5]$. Otherwise, if $x\in[-1,1]$, our implementation also requires an evaluation of the square root. For evaluating inverse square roots, we optimize the initial guess which was also used in~\cite{Bhaskar:2016:QAC:3179448.3179450} in order to reduce the number of required Newton iterations by 1 (which corresponds to a reduction by $20$-$25\%$). In contrast to the mentioned works, we implement all our high-level arithmetic functions at the level of Toffoli gates in the quantum programming language LIQ$Ui\Ket{}$. As a result, we were able to test our circuits on various test vectors using a Toffoli circuit simulator, ranging up to several hundreds of qubits.

Throughout this paper, we adapt ideas from classical high-performance computing in order to reduce the required resources in the quantum setting. While these methods allow to reduce the Toffoli and qubit counts significantly, the resulting circuits are still quite expensive, especially in terms of the number of gates that are required. We hope that this highlights the fact that more research in the implementation of quantum algorithms is necessary in order to further reduce the cost originating from arithmetic in the computational basis.

\section{Learning from Classical Arithmetic Libraries}
While there is no need for computations to be reversible when using classical computers, a significant overlap of techniques from reversible computing can be found in vectorized high-performance libraries. In quantum computing, having an if-statement collapses the state vector, resulting in a loss of all potential speedup. Similarly, if-statements in vectorized code require a read-out of the vector, followed by a case distinction and a read-in of the handled values, which incurs a tremendous overhead and results in a deterioration of the expected speedup or even an overall slowdown. Analogous considerations have to be taken into account when dealing with, e.g., loops.
Therefore, classical high-performance libraries may offer ideas and insights applicable to quantum computing, especially for mathematical functions such as (inverse) trigonometric functions, exponentials, logarithms, etc., of which highly-optimized implementations are available in, e.g., the Cephes math library~\cite{moshier2000cephes} or games such as Quake III Arena (their fast inverse square root~\cite{quake3} is reviewed in~\cite{lomont2003fast}).

Although some of these implementations rely on a floating-point representation, many ideas carry over to the fixed-point domain, and remain efficient enough even when requiring reversibility. Specifically, we adapt implementations of the arcsine function from~\cite{moshier2000cephes} and the fast inverse square root from~\cite{lomont2003fast} to the quantum domain by providing reversible low-level implementations. Furthermore, we describe a parallel version of the classical Horner scheme~\cite{knuth1962evaluation}, which enables the conditional evaluation of many polynomials in parallel and, therefore, efficient evaluation of piecewise polynomial approximations.

\begin{figure*}[t]
	\includegraphics[width=\linewidth]{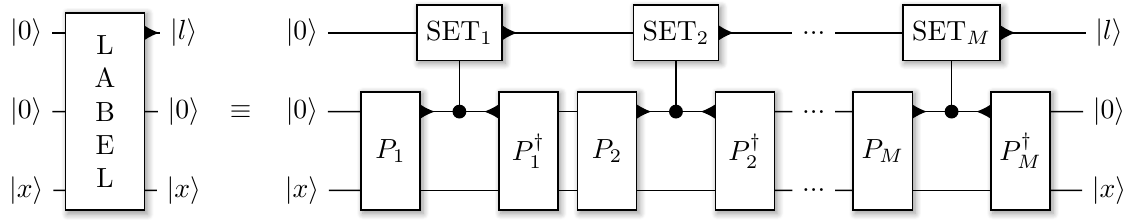}
	\caption{The LABEL gate initializes the label register $\Ket l$, which consists of $\lceil\log_2(M)\rceil$ qubits, to indicate the subdomain $\Omega_l$ to which $x$ belongs. $P_i$ computes the predicate indicating whether $x\in\Omega_i$ into the ancilla qubit. Conditioned on this result, the label is then initialized to the value chosen to represent the $i$-th interval.}
	\label{fig:label}
\end{figure*}

\begin{figure}[t]
	\includegraphics[width=\linewidth]{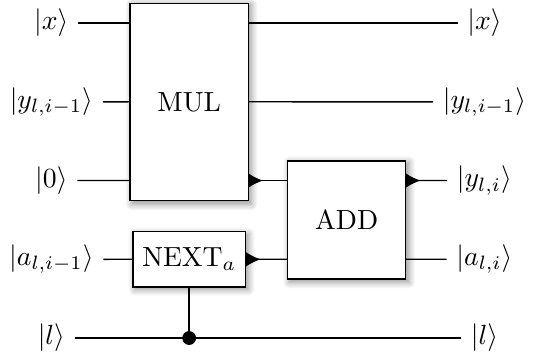}
	\caption{Our parallel polynomial evaluation circuit. NEXT$_a$ changes the register to hold the next set of coefficients (in superposition) $\sum_l\Ket l\Ket{a_{l,i-1}}\mapsto \sum_l\Ket l\Ket{a_{l,i}}$. MUL and ADD perform a multiplication and an addition, respectively. The small triangle indicates the output of the ADD and MUL gates.}
	\label{fig:ppoly}
\end{figure}

\section{Evaluation of piecewise polynomial approximations}

A basic scheme to evaluate a single polynomial on a quantum computer in the computational basis is the classical Horner scheme, which evaluates
\[
	P(x) = \sum_{i=0}^d a_ix^i
\]
by iteratively performing a multiplication by $x$, followed by an addition of $a_i$ for $i\in\{d,d-1,...,0\}$. This amounts to performing the following operations:
\begin{align*}
	a_d x + a_{d-1}&\mapsto a_dx^2+a_{d-1}x+a_{d-2}\\
	&\cdots\\
	&\mapsto a_dx^d+\cdots+a_0\;.
\end{align*}

A reversible implementation of this scheme simply stores all intermediate results. At iteration $i$, the last iterate $y_{i-1}$ is multiplied by $x$ into a new register $y_{i}$, followed by an addition by the (classically-known) constant $a_i$, which may make use, e.g., the addition circuit by Takahashi~\cite{takahashi2009quantum} (if there is an extra register left), or the in-place constant adder by H\"aner et al.~\cite{haner2016factoring}, which does not require an ancilla register but is more costly in terms of gates.
Due to the linear dependence of successive iterates, a pebbling strategy can be employed in order to optimize the space/time trade-offs according to some chosen metric~\cite{parent2015reversible}.

Oftentimes, the degree $d$ of the minimax approximation over a domain $\Omega$ must be chosen to be very high in order to achieve a certain $L_\infty(\Omega)$-error. In such cases, it makes sense to partition $\Omega$, i.e., find $\Omega_i$ such that
\[
	\Omega = \bigcup_{i=0}^M \Omega_i\;,\;\Omega_i\cap\Omega_j=\emptyset\;\forall i\neq j\;,
\]
and to then perform a case distinction for each input, evaluating a different polynomial for $x\in\Omega_i$ than for $y\in\Omega_j$ if $i\neq j$. A straight-forward generalization of this approach to the realm of quantum computing would loop over all subdomains $\Omega_i$ and, conditioned on a case-distinction or label register $\Ket l$, evaluate the corresponding polynomial. Thus, the cost of this inefficient approach grows linearly with the number of subdomains.

In order to improve upon this approach, one can parallelize the polynomial evaluation if the degree $d$ is constant over the entire domain $\Omega$. Note that merely adding the label register $\Ket l$ mentioned above and performing
\begin{align}\label{eqn:init2}
	\Ket{y_{l,i-1}x}\Ket0\Ket l&\mapsto \Ket{y_{l,i-1}x}\Ket{a_{l,i}}\Ket l\\
	&\mapsto \Ket{y_{l,i-1}x+a_{l,i}}\Ket{a_{l,i}}\Ket l\\
	&\mapsto\Ket{y_{l,i}}\Ket0\Ket l\;,
\end{align}
enables the evaluation of multiple polynomials in parallel. The impact on the circuit size is minor, as will be shown in Appendix~\ref{sec:polycost}. The depth of the circuit remains virtually unaltered, since the initialization step \eqref{eqn:init2} can be performed while multiplying the previous iterate $y_{i-1}$ by $x$, see Fig.~\ref{fig:ppoly}. An illustration of the circuit computing the label register $\Ket l$ can be found in Fig.~\ref{fig:label}. A slight drawback of this parallel evaluation is that it requires one extra ancilla register for the last iteration, since the in-place addition circuit~\cite{haner2016factoring} can no longer be used. Resource estimates of a few functions which were implemented using this approach can be found in Table~\ref{tbl:funcs}. The small overhead of using many intervals allows to achieve good approximations already for low-degree polynomials (and thus using few qubit registers).

Using reversible pebble games \cite{Bennett:89}, it is possible to trade the number of registers needed to store the iterates with the depth of the resulting circuit. The parameters are: the number $n$ of bits per register, the total number $m$ of these $n$-qubit registers, the number $r$ of Horner iterations, and the depth $d$ of the resulting circuit. The trade-space we consider involves $m$, $r$, and $d$. In particular, we consider the question of what the optimal circuit depth is for a fixed number $m$ of registers and a fixed number $r$ of iterations. 
As in \cite{Knill:95,PRS:2015} we use dynamic programming to construct the optimal strategies as the dependency graph is just a line which is due to the sequential nature of Horner's method (the general pebbling problem is much harder to solve, in fact finding the optimal strategy for general graphs is known to be PSPACE complete \cite{Chan:2013}). The optimal number of pebbling steps as a function of $m$ and $r$ can be found in Table~\ref{tab:pebbling}.

\begin{table}
\centering
\begin{tabular}{c@{\qquad}c@{\;\;}c@{\;\;}c@{\;\;}c@{\;\;}c@{\;\;}c@{\;\;}c@{\;\;}c@{\qquad}c@{\qquad}c@{\qquad}c}
\hline\hline \\[-2ex]
$m \backslash r$ & 1 & 2 & 3 & 4 & 5 & 6 & 7 & 8 & 16 & 32 & 64 \\
\hline 
1 & 1 &$\infty$ &$\infty$ &$\infty$ &$\infty$ &$\infty$ &$\infty$ &$\infty$ &$\infty$ &$\infty$ &$\infty$ \\
2 & 1 &     3 &$\infty$ &$\infty$ &$\infty$ &$\infty$ &$\infty$ &$\infty$ &$\infty$ &$\infty$ &$\infty$ \\
3 & 1 &     3 &     5 &     9 &$\infty$ &$\infty$ &$\infty$ &$\infty$ &$\infty$ &$\infty$ &$\infty$ \\
4 & 1 &     3 &     5 &     7 &    11 &    15 &    19 &    25 &$\infty$ &$\infty$ &$\infty$ \\
5 & 1 &     3 &     5 &     7 &     9 &    13 &    17 &    21 &    71 &$\infty$ &$\infty$ \\
6 & 1 &     3 &     5 &     7 &     9 &    11 &    15 &    19 &    51 &   193 &$\infty$ \\
7 & 1 &     3 &     5 &     7 &     9 &    11 &    13 &    17 &    49 &   145 &   531 \\
8 & 1 &     3 &     5 &     7 &     9 &    11 &    13 &    15 &    47 &   117 &   369 \\
\hline\hline
\end{tabular}
\caption{\label{tab:pebbling} Optimal pebbling strategies for fixed number $m$ of registers and fixed number $r$ of sequential iterations. This table can be used for both the Horner scheme for polynomial evaluation, where $r$ corresponds to the polynomial degree, and for Newton's method, where $r$ denotes the number of iterations. The number for entry $(m,r)$ denotes how many pebbling steps it takes to compute the entire sequence. The circuit width and depth can be obtained from these numbers.}
\end{table}

\section{Software Stack Module for piecewise smooth functions}

In order to enable automatic compilation of an oracle which implements a piecewise smooth function, the Remez algorithm~\cite{remez1934determination} can be used in a subroutine to determine a piecewise polynomial approximation, which can then be implemented using the circuit described in the previous section.

In particular, we aim to implement the oracle with a given precision, accuracy, and number of available quantum registers (or, equivalently, the polynomial degree $d$ if no pebbling is employed) over a user-specified interval $\Omega=[a,a+L)$. Our algorithm proceeds as follows: In a first step, run the Remez algorithm which, given a function $f(x)$ over a domain $\Omega\subset\mathbb R$ and a polynomial degree $d$, finds the polynomial $P(x)$ which approximates $f(x)$ with minimal $L_\infty(\Omega)$-error, and check whether the achieved error is low enough. If it is too large, reduce the size of the domain $\Omega_1:=[a,a+\frac L2)$ and check again. Repeating this procedure and carrying out binary search on the right interval border will eventually lead to the first subdomain $\Omega_1=[a,b_1)$ which is the largest interval such that the corresponding degree $d$ polynomial achieves the desired accuracy. Next, one determines the next subdomain $\Omega_2=[b_1,b_2)$ using the same procedure. This is iterated until $b_i\geq b$, meaning that all required subdomains and their corresponding polynomials have been determined and $f(x)$ can be implemented using a parallel polynomial evaluation circuit. This algorithm was implemented and then run for various functions, target accuracies, and polynomial degrees in order to determine approximate resource estimates for these parameters, see Table~\ref{tbl:funcs} in the appendix.

\begin{figure}[!t]
	\subfigure[Before constant-tuning.]{
	\resizebox{.95\linewidth}{!}{\input{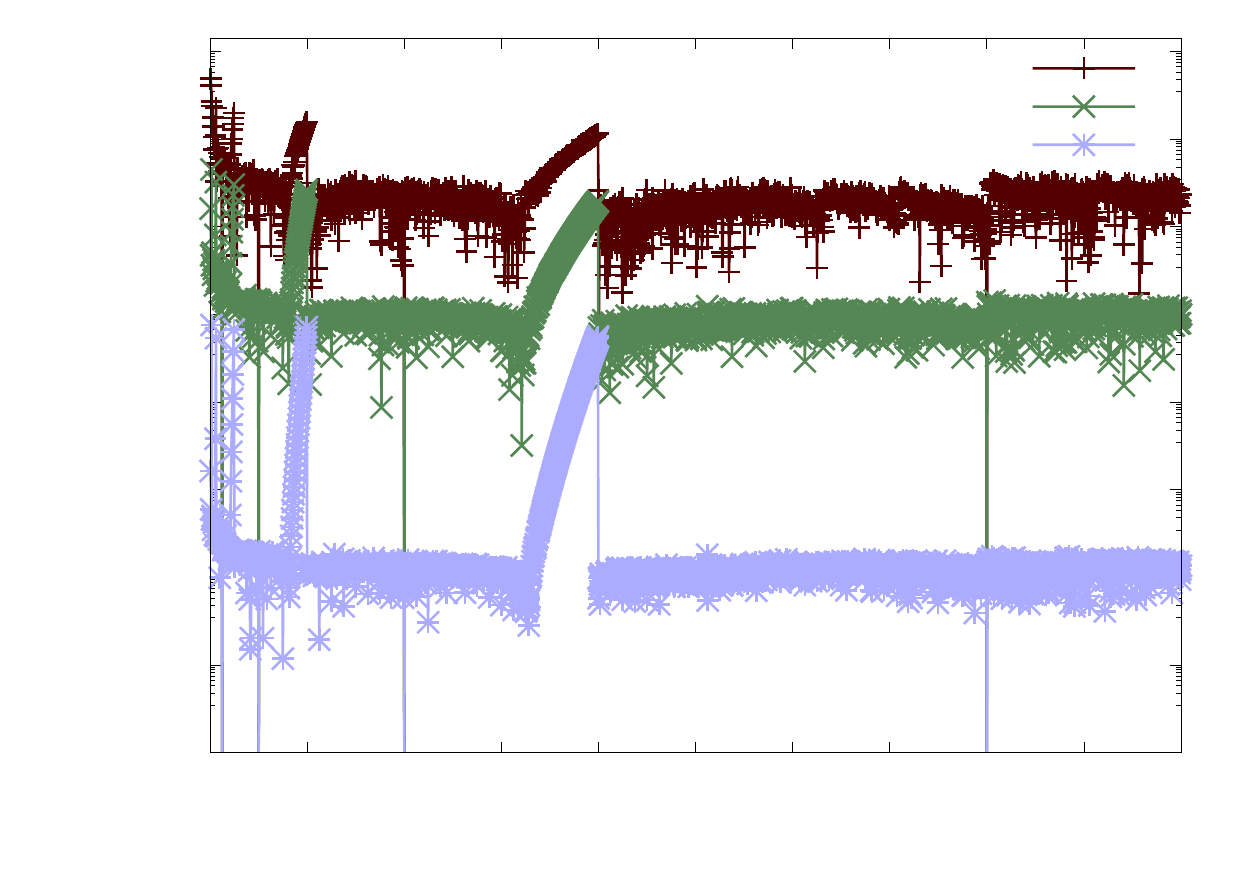}}
	\label{fig:invsqrterrornoopt}
	}
	\subfigure[After constant-tuning.]{
	\resizebox{.95\linewidth}{!}{\input{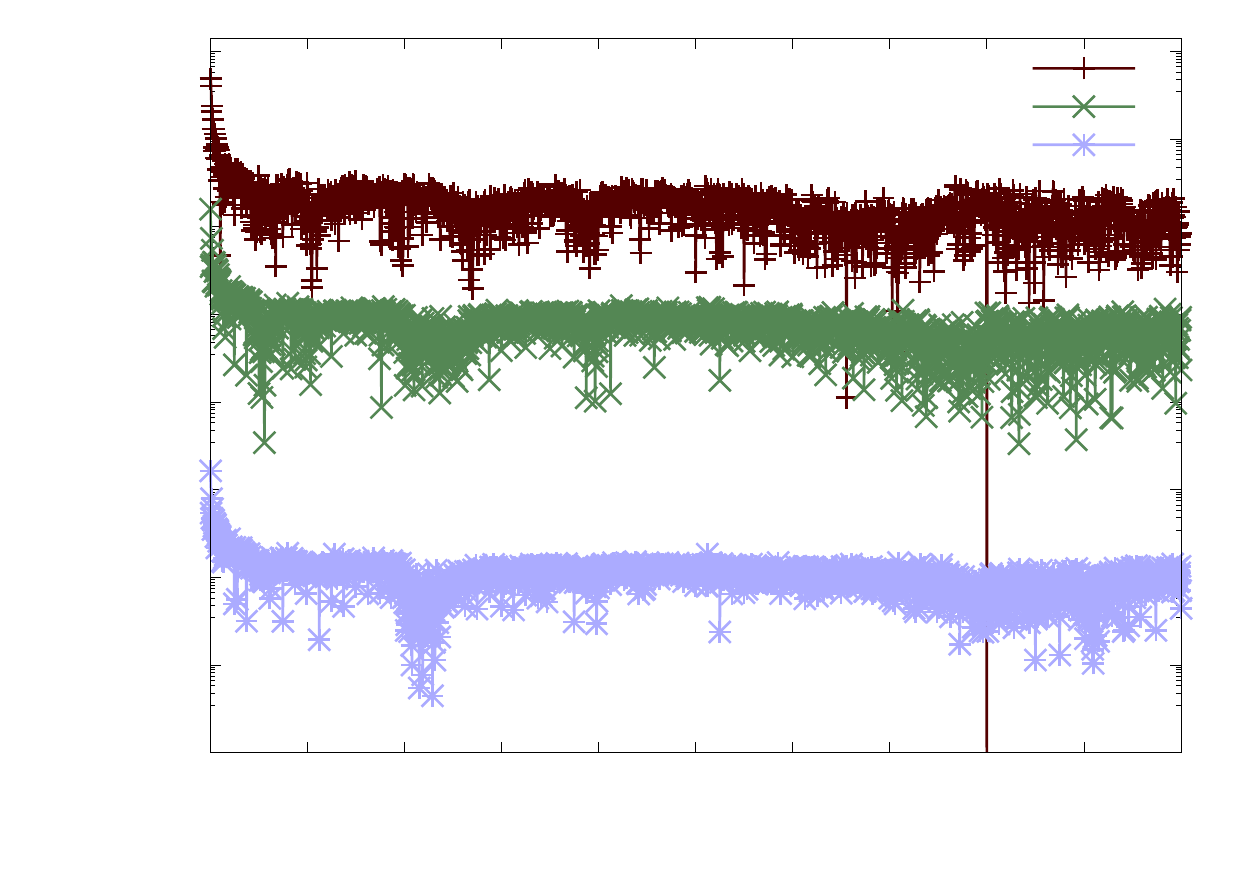}}
	\label{fig:invsqrterror}
	}
	\caption{Absolute errors of the inverse square root before and after tuning the constants (see Eqn.~\ref{eq:initguessconstant}). The errors were evaluated for $N=2000$ (equidistant) points in the interval $[\frac 1N,5]$ using $m\in\{2,3,4\}$ Newton iterations and corresponding bit sizes $n\in\{25,35,55\}$. The fixed-point position is $p=12$, in order to ensure that no overflow occurs for small inputs.}
\end{figure}

\section{Inverse square root}

For quantum chemistry or machine learning applications, also non-smooth functions are required. Most notably, the inverse square root can be used in both examples, namely for the calculation of the Coulomb potential and to determine the reciprocal when employing HHL~\cite{harrow2009quantum} for quantum machine learning.

In classical computing, inverse square roots appear in computer graphics and the term ``fast inverse square root'' is often used: It labels the procedure to approximate the inverse square root using bit-operations on the floating-point representation of the input, as it was done in Quake III Arena~\cite{quake3} (see~\cite{lomont2003fast} for a review). The code ultimately performs a Newton-Raphson iteration in order to improve upon a pretty accurate initial guess, which it finds using afore-mentioned bit-operations. Loosely speaking, the bit-operations consist of a bit-shift to divide the exponent by two in order to approximate the square root, followed by a subtraction of this result from a \textit{magic number}, effectively negating the exponent and correcting the mantissa, which was also shifted together with the exponent. The \textit{magic number} can be chosen using an auto-tuning procedure and varies depending on the objective function being used~\cite{lomont2003fast}. This provides an extremely good initial guess for the Newton iteration at very low cost.

In our reversible implementation, we use a similar procedure to compute the inverse square root using fixed-point arithmetic. While we cannot make use of the floating-point representation, we can still find a low-cost initial guess which allows for a small number of Newton iterations to be sufficient (i.e., 2-4 iterations). This includes determining the position of the first one in the bit-representation of the input, followed by an initialization which involves a case distinction on the \textit{magic number} to use. Our three magic constants (see Appendix~\ref{sec:newton}) were tuned such that the error peaks near powers of two in Fig.~\ref{fig:invsqrterrornoopt} vanish. The peaks appear due to the fact that the initial guess takes into account the location of the first one but completely ignores the actual magnitude of the input. For example, all inputs in $[1,2)$ yield the same initial guess. The error plot with tuned constants is depicted in Fig.~\ref{fig:invsqrterror}. One can clearly observe that an entire Newton iteration can be saved when aiming for a given $L_\infty$-error.

\begin{figure}[t]
	\resizebox{.95\linewidth}{!}{\input{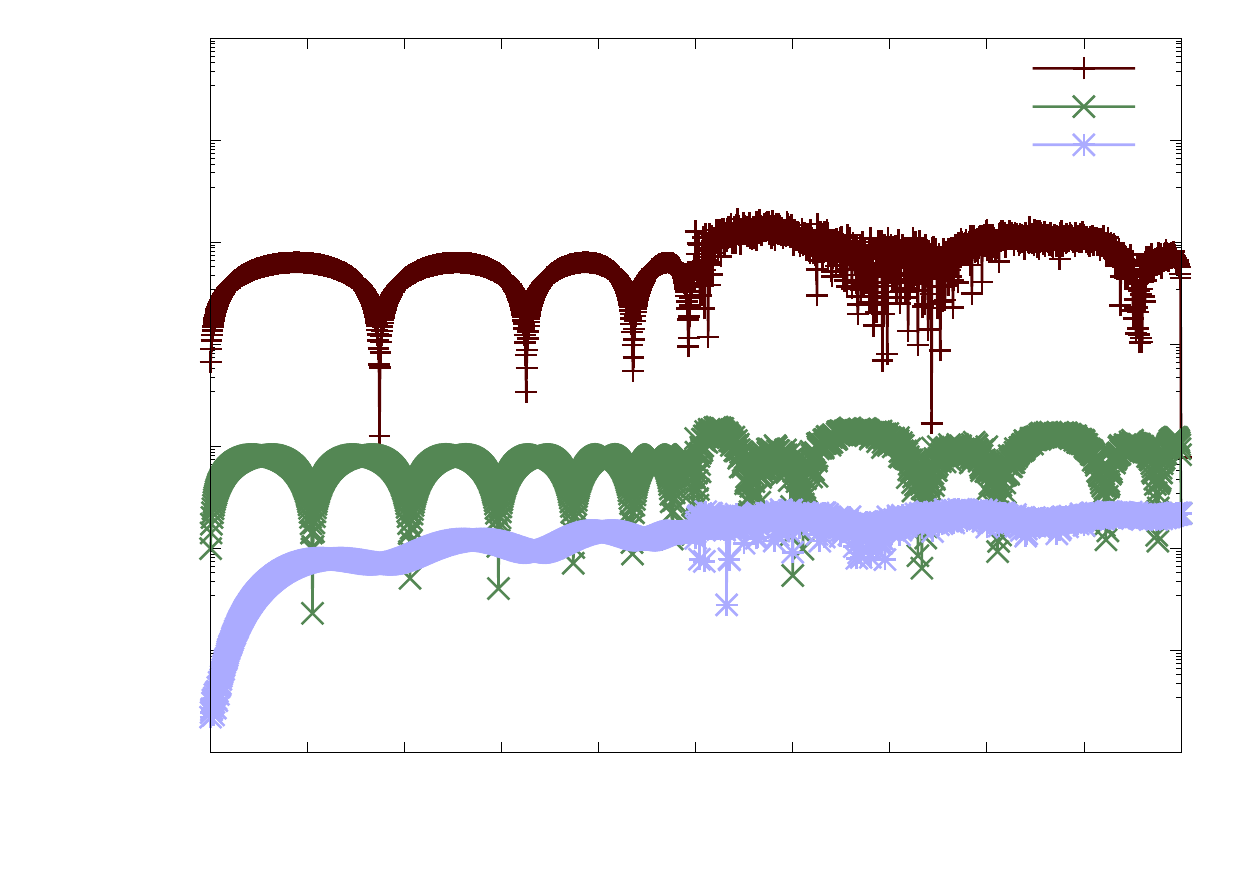}}
	\caption{Absolute error on $[0,1]$ for $N=2000$ points of our reversible implementation of the arcsine using $m\in\{3,4,5\}$ Newton iterations for calculating the inverse square root. The fixed-point position is chosen to be $p=2$ and total bit size $n$ was chosen to be in $\{35,50,55\}$.}
	\label{fig:arcsinerror}
\end{figure}

\section{Arcsine}

Following the implementation used in the classical math library Cephes~\cite{moshier2000cephes}, an arcsine can be implemented as a combination of polynomial evaluation and the square root. Approximating the arcsine using only a polynomial allows for a good approximation in $[-0.5,0.5]$, but not near $\pm1$ (where it diverges). The Cephes math library remedies this problem by adding a case distinction, employing a ``double-angle identity'' for $|x|\geq 0.5$. This requires computing the square root, which can be achieved by first calculating the inverse square root, followed by $x\cdot\frac 1{\sqrt x}=\sqrt x$. Alternatively, the new square root circuit from Ref.~\cite{MT:2018} can be used.

We have implemented our circuit for arcsine and we show the resulting error plot in Fig.~\ref{fig:arcsinerror}. The oscillations stem from the minimax polynomial which is used to approximate the arcsine on $[-0.5,0.5]$.
More implementation details and resource estimates can be found in Appendix~\ref{sec:arcsine}.

Note that certain applications may allow to trade off error in the arcsine with, e.g., probability of success by rescaling the input such that the arcsine needs to be computed only for values in $[-0.5,0.5]$. This would allow one to remove the case-distinction and the subsequent calculation of the square root: One could evaluate the arcsine at a cost that is similar to the implementation costs of sin/cos. Estimates for the Toffoli and qubit counts for this case can also be found in the appendix, see Table~\ref{tbl:funcs}.

\section{Summary and Outlook}
We have presented efficient quantum circuits for the evaluation of many mathematical functions, including (inverse) square root, Gaussians, hyperbolic tangent, exponential, sine/cosine, and arcsine. Our circuits can be used to obtain accurate resource estimates for various quantum algorithms and the results may help to identify the first large-scale applications as well as bottlenecks in these algorithms where more research is necessary in order to make the resource requirements practical. When embedded in a quantum compilation framework, our general parallel polynomial evaluation circuit can be used for automatic code generation when compiling oracles that compute piecewise smooth mathematical functions in the computational basis. This tremendously facilitates the implementation of quantum algorithms which employ oracles that compute such functions on a superposition of inputs.

\bibliographystyle{unsrt}
\bibliography{references}

\appendix

\section{Basic circuit building blocks for fixed-point arithmetic}\label{sec:basiccircuits}
In fixed-point arithmetic, one represents numbers $x$ using $n$ bits as
\[
	x=\underbrace{x_{n-1}\cdots x_{n-p}}_p.\underbrace{x_{n-p-1}\cdots x_0}_{n-p}\;,
\]
where $x_i\in\{0,1\}$ is the $i$-th bit of the binary representation of $x$, and the point position $p$ denotes the number of binary digits to the left of the binary point. We choose both the total number of bits $n$ and the point position $p$ to be constant over the course of a computation. As a consequence, over- and underflow errors are introduced, while keeping the required bit-size from growing with each operation.

\paragraph{Fixed-point addition.} We use a fixed-point addition implementation, which keeps the bit-size constant. This amounts to allowing over- and underflow, while keeping the registers from growing with each operation.

\label{sec:mult}
\paragraph{Fixed-point multiplication.} Multiplication can be performed by repeated-addition-and-shift, which can be seen from
\[
	x\cdot y=x_{n-1}2^{n-1} y+\cdots+x_02^0y\;,
\]
where $x=\sum_i x_i2^i$ with $x_i\in\{0,1\}$ denotes the binary expansion of the $n$-bit number $x$. Thus, for $i\in\{0,...,n-1\}$, $2^{i-(n-p)}y$ is added to the result register (which is initially zero) if $x_i=1$. This can be implemented using $n$ controlled additions on $1,2,...,n$ bits if one allows for pre-truncation: Instead of computing the $2n$-bit result and copying out the first $n$ bits before uncomputing the multiplication again, the additions can be executed on a subset of the qubits, ignoring all bits beyond the scope of the $n$-bit result. Thus, each addition introduces an error of at most $\varepsilon_A=\frac 1{2^{n-p}}$. Since there are (at most) $n$ such additions, the total error is
\[
	\varepsilon=\frac n{2^{n-p}}\;,
\]
a factor $n$ larger than using the costly approach mentioned above.

Negative multipliers are dealt with by substituting the controlled addition by a controlled subtraction when conditioning on the most significant bit~\cite{wakerly2000digital} because it has negative weight $w_{MSB}=-2^{n-1}$ in two's-complement notation.
The multiplicand is assumed to be positive throughout, which removes the need for conditional inversions of input and output (for every multiplication), thus tremendously reducing the size of circuits that require many multiplications such as, e.g., polynomial evaluation.

\paragraph{Fixed-point squaring.} The square of a number can be calculated using the same approach as for multiplication. Yet, one can save (almost) an entire register by only copying out the bit being conditioned on prior to performing the controlled addition. Then, the bit can be reset using another CNOT gate, followed by copying out the next bit and performing the next controlled addition. The gate counts are identical to performing
\[
	\Ket x\Ket0\Ket0\mapsto\Ket x\Ket x\Ket 0\mapsto \Ket x\Ket x\Ket{x^2}\mapsto\Ket x\Ket{x^2}\Ket 0\;,
\]
while allowing to save $n-1$ qubits.

\section{Resource estimates for polynomial evaluation}
\label{sec:polycost}

The evaluation of a degree $d$ polynomial requires an initial multiplication $a_d \cdot x$, an addition of $a_{d-1}$, followed by $d-1$ multiply-accumulate instructions.
The total number of Toffoli gates is thus equal to the cost of $d$ multiply-accumulate instructions. Furthermore, $d+1$ registers are required for holding intermediate and final result(s) if no in-place adder is used for the last iteration (and no non-trivial pebbling strategy is applied). Other strategies may be employed in order to reduce the number of ancilla registers, at the cost of a larger gate count, see Table~\ref{tab:pebbling} for examples.

Note that all multiplications can be carried out assuming $x>0$, i.e. $x$ can be conditionally inverted prior to the polynomial evaluation (and the pseudo-sign bit is copied out). The sign is then absorbed into the coefficients: Before adding $a_i$ into the $\Ket{y_{i-1}x}$-register, it is inverted conditioned on the sign-bit of $x$ being set if the coefficient corresponds to an odd power. This is done because it is cheaper to implement a fixed-point multiplier which can only deal with $y_{i-1}$ being negative (see Sec.~\ref{sec:basiccircuits}).

The Toffoli gate count of multiplying two $n$-bit numbers (using truncated additions as described in Sec.~\ref{sec:mult}) is
\begin{align*}
T_\text{mul}(n,p)&=\sum_{i=0}^{p-1} T_\text{cadd}(n-i)+\sum_{i=1}^{n-p}T_\text{cadd}(n-i)\\
&=\sum_{i=0}^{p-1} 3(n-i)+\sum_{i=1}^{n-p}3(n-i)+3n\\
&=\frac 32n^2 + 3np+\frac 32 n - 3p^2+3p
\end{align*}
if one uses the controlled addition circuit by Takahashi et al.~\cite{takahashi2009quantum}, which requires $3n+3$ Toffoli gates to (conditionally) add two $n$-bit numbers. The subsequent addition can be implemented using the addition circuit by Takahashi et al.~\cite{takahashi2009quantum}, featuring $2n-1$ Toffoli gates. Thus, the total cost of a fused multiply-accumulate instruction is
\[
	T_\text{fma}(n,p)=\frac 32n^2 + 3np+\frac{7}2 n - 3p^2+3p-1\;.
\]
Therefore, the total Toffoli count for evaluating a degree $d$ polynomial is
\[
	T_\text{poly}(n,d,p)=\frac 32n^2d + 3npd+\frac{7}2 nd - 3p^2d+3pd-d\;.
\]

Evaluating $M$ polynomials in parallel for piecewise polynomial approximation requires only $n+\lceil\log_2 M\rceil$ additional qubits (since one $n$-qubit register is required to perform the addition in the last iteration, which is no longer just a constant) and $2M$ $\lceil\log_2M\rceil$-controlled NOT gates, which can be performed in parallel with the multiplication. This increases the circuit size by 
\[
	T_\text{extra}(M)=2M(4\lceil\log_2M\rceil - 8)
\]
Toffoli gates per multiply-accumulate instruction, since a $k$-controlled NOT can be achieved using $4(k-2)$ Toffoli gates and $k-2$ dirty ancilla qubits~\cite{barenco1995elementary}, which are readily available in this construction.

The label register $\Ket l$ can be computed using 1 comparator per subinterval
\[
	I_i = [a_i,a_{i_{i+1}}),\;a_0<a_1<...<a_{M-1}\;.
\]
The comparator stores its output into one extra qubit, flipping it to $\Ket 1$ if $x \leq a_{i+1}$. The label register is then incremented from $i-1$ to $i$, conditioned on this output qubit still being $\Ket 0$ (indicating that $x > a_i$). Incrementing $\Ket l$ can be achieved using CNOT gates applied to the qubits that correspond to ones in the bit-representation of $(i-1)\oplus i$. Finally, the comparator output qubit is uncomputed again. This procedure is carried out $M$ times for $i=0,...,M-1$ and requires 1 additional qubit. The number of extra Toffoli gates for this label initialization is
\begin{align*}
	T_\text{label}(M,n) &= M\cdot 2T_\text{cmp}(n)\\
	&=4Mn\;,
\end{align*}
where, as a comparator, we use the CARRY-circuit from~\cite{haner2016factoring}, which needs $2n$ Toffoli gates to compare a classical value to a quantum register, and another $2n$ to uncompute the output and intermediate changes to the $n$ required dirty ancilla qubits.

In total, the parallel polynomial evaluation circuit thus requires
\begin{align*}
	T_\text{pp}(n,d,p,M)&=T_\text{poly}(n,d,p)+d\cdot T_\text{extra}(M)\\
	&\phantom{={}}+T_\text{label}(M,n)\\
	&=\frac 32n^2d + 3npd+\frac{7}2 nd - 3p^2d+3pd-d\\
	&\phantom{={}}+2Md(4\lceil\log_2M\rceil-8)+4Mn
\end{align*}
Toffoli gates and $(d+1)n+\lceil\log_2M\rceil + 1$ qubits.

\section{(Inverse) Square root}\label{sec:newton}

The inverse square root, i.e.,
\[
	f(x)=\frac 1{\sqrt x}
\]
can be computed efficiently using Newton's method. The iteration looks as follows:
\[
	x_{n+1} = x_n\left(1.5-\frac{ax_n^2}2\right)\;,
\]
where $a$ is the input and $x_n\overset{n\rightarrow\infty}{\longrightarrow} \frac 1{\sqrt a}$ if the initial guess is sufficiently close to the true solution.
\subsection{Reversible implementation}
\paragraph{Initial guess and first round.}
Finding a good initial guess $x_0\approx \frac 1{\sqrt a}$ for Newton's zero-finding routine is crucial for (fast) convergence. A crude approximation which turns out to be sufficient is the following:
\[
	\frac 1{\sqrt a} = \left(2^{\log_2 a}\right)^{-\frac 12}=2^{-\frac{\log_2a}2}\approx 2^{\lfloor -\frac{\lfloor\log_2a\rfloor}2\rceil}=\tilde x_0\;,
\]
where $\lfloor\log_2 a\rfloor$ can be determined by finding the first ``1'' when traversing the bit-representation of $a$ from left to right (MSB to LSB).
While the space requirement for $\tilde x_0$ is in $\mathcal O(\log_2 n)$, such a representation would be impractical for the first Newton round. Furthermore, noting that the first iteration on $\tilde x_0=2^k$ leads to
\begin{equation}\label{eq:initialguess}
	\tilde x_1 = 2^k\left(1.5-\frac{a2^{2k}}2\right)=:x_0\;,
\end{equation}
one can directly choose this $x_0$ as the initial guess. The preparation of $x_0$ can be achieved using $(n-1)+n+1$ ancilla qubits, which must be available due to the space requirements of the subsequent Newton steps. The one ancilla qubit is used as a flag indicating whether the first ``1'' from the left has already been encountered. For each iteration $i\in \{n-1,...,1,0\}$, one determines whether the bit $a_{i}$ is 1 and stores this result $r_i$ in one of the $n$ work qubits, conditioned on the flag being unset. Then, conditioned on $r_i=1$, the flag is flipped, indicating that the first ``1'' has been found. If $r_i=1$, the $x_0$-register is initialized to the value in \eqref{eq:initialguess} as follows: Using CNOTs, the $x_0$-register can be initialized to the value $1.5$ shifted by $k=\frac{p-2i}2$, where $p$ denotes the binary point position of the input, followed by subtracting the $(3k-1)$-shifted input $a$ from $x_0$, which may require up to $n-1$ ancilla qubits.

In order to improve the quality of the first guess for numbers close to $2^k$ for some $k\in\mathbb Z$, one can tune the constant $1.5$ in $\eqref{eq:initialguess}$, i.e., turn it into a function $C(k)$ of the exponent $k$. This increases the overall cost of calculating $x_0$ merely by a few CNOT gates but allows to save an entire Newton iteration even when only distinguishing three cases, namely
\begin{equation}\label{eq:initguessconstant}
	C(k):=\left\{\begin{matrix} 1.613, & k < 0\\
	1.5,& k = 0\\
	1.62,& k > 0\end{matrix}\right.\;.
\end{equation}

\begin{figure}[ht]
	\centering
	\includegraphics[width=\linewidth]{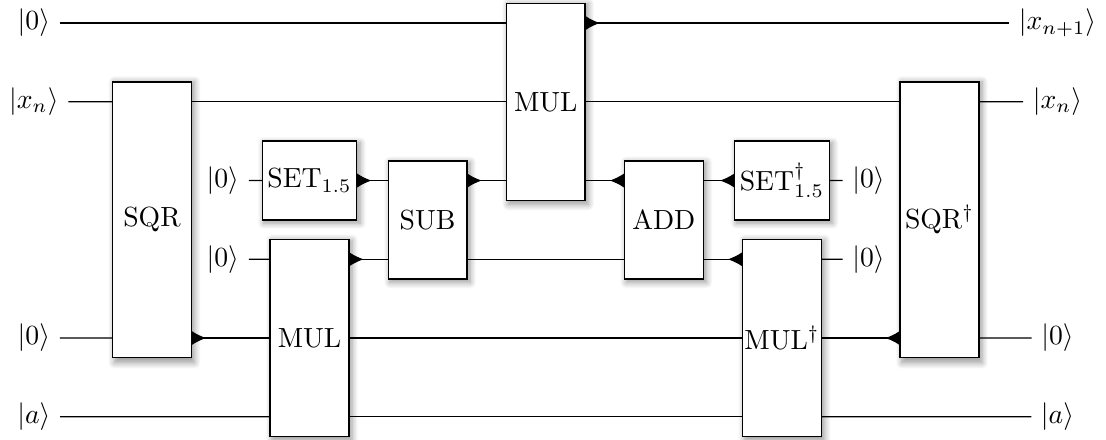}
	\caption{Circuit for the $n$-th Newton iteration of computing the inverse square root of $a$, given in a quantum superposition in $\Ket a$. SQR computes the square of the previous iterate $x_n$ into an empty result-register, which is then multiplied by the input $a$ (MUL), followed by subtracting (SUB) this intermediate result from the value $1.5$ (initialized using the SET$_{1.5}$-gate). Finally, the next iterate, i.e., $x_{n+1}=x_n(1.5-\frac 12ax_n^2)$ can be computed by multiplying this intermediate result by $x_n$. All temporary results are then cleared by running the appropriate operations in reverse order.}
	\label{fig:newtonit}
\end{figure}

\paragraph{The Newton iteration.}
Computing $x_{n+1}$ from $x_n$ by
\[
	x_{n+1} = x_n\left(1.5 - \frac{ax_n^2}2\right)\;,
\]
can be achieved as follows:

\begin{enumerate}[1.]
	\item Compute the square of $x_n$ into a new register.
	\item Multiply $x_n^2$ by the shifted input to obtain $ax_n^2/2$.
	\item Initialize another register to 1.5 and subtract $ax_n^2/2$.
	\item Multiply the result by $x_n$ to arrive at $x_{n+1}$.
	\item Uncompute the three intermediate results.
\end{enumerate}
The circuit of one such Newton iteration is depicted in Fig.~\ref{fig:newtonit}.

Therefore, for $m$ Newton iterations, this requires $m+3$ $n$-qubit registers if no pebbling is done on the Newton iterates, i.e., if all $x_i$ are kept in memory until the last Newton iteration has been completed.

\subsection{Resource estimates}
Computing the initial guess for the fast inverse square root requires $n$ controlled additions of two $n$-bit numbers plus $2n$ Toffoli gates for checking/setting the flag (and uncomputing it again). Thus, the Toffoli count for the initial guess is
\[
	T_\text{init}(n)=nT_\text{cadd}(n)+2n=3n^2+5n\;.
\]
Each Newton iteration features squaring, a multiplication, a subtraction, a final multiplication (yielding the next iterate), and then an uncomputation of the three intermediate results. In total, one thus employs 5 multiplications and 2 additions (of which 2 multiplications and 1 addition are run in reverse), which yields the Toffoli count
\begin{align*}
	T_\text{iter}(n,p)&=5T_\text{mul}(n,p)+2T_\text{add}(n)\\
	&=\frac{15}2 n^2+15 n p+\frac{23}2 n-15 p^2+15 p-2\;.
\end{align*}
The number of Toffoli gates for the entire Newton procedure (without uncomputing the iterates) for $m$ iterations thus reads
\begin{align*}
	T_\text{invsqrt}(n,m,p)&=T_\text{init}(n)+mT_\text{iter}(n,p)\\
	&= n^2(\frac{15}2 m+3)+15 n p m+ n (\frac{23}2m+5)\\
	&\phantom{={}}-15 p^2 m+15 p m-2m\;.
\end{align*}
Since each Newton iteration requires $3$ ancilla registers (which are cleaned up after each round) to produce the next iterate, the total number of qubits is $n(m+4)$, where one register holds the initial guess $x_0$.

Note that this is an upperbound on the required number of both qubits and Toffoli gates. Since Newton converges quadratically, there is no need to perform full additions and multiplications at each iteration. Rather, the number of bits $n$ used for the fixed point representation should be an (increasing) function of the Newton iteration.

The square root can be calculated using
\[
	\sqrt{x} = x\cdot\frac 1{\sqrt x}\;,
\]
i.e., at a cost of an additional multiplication into a new register. Note that this new register would be required anyway when copying out the result and running the entire computation in reverse, in order to clear registers holding intermediate results. Thus, the total number of logical qubits remains unchanged.

\section{Arcsine}\label{sec:arcsine}
While $\sin(x)$ and $\cos(x)$ are very easy to approximate using, e.g., polynomials, their inverses are not. The main difficulty arises near $\pm 1$, where 
\[
	\frac {d\arcsin(x)}{dx}=\frac 1{\sqrt{1-x^2}}
\]
diverges. Therefore, it makes sense to use an alternative representation of $\arcsin(x)$ for larger values of $x$, e.g.,
\begin{align*}
	\arcsin(x) &= \frac{\pi}{2} - \arccos(x)\\
	&=\frac{\pi}{2} - \arcsin\left(\sqrt{1-x^2}\right)\;.
\end{align*}
Applying the double-argument identity to the last expression yields
\begin{equation}\label{eqn:arcsin}
	\arcsin(x) = \frac{\pi}{2} - 2\arcsin\left(\sqrt{\frac{1-x}2}\right)\;,
\end{equation}
a very useful identity which was already used in a classical math library called Cephes~\cite{moshier2000cephes}. We use the same partitioning of the interval, using a minimax polynomial to approximate $\arcsin(x)$ for $x\in[0,0.5)$, and the transformation in \eqref{eqn:arcsin} for $x\in[0.5,1]$. We use our inverse square root implementation to compute $\sqrt{z}$ for
\[
	z = \frac{1-x}2\;,
\]
which satisfies $z\in[0,0.25]$, for $x\in[0.5,1]$. Therefore, the fixed point position has to be chosen large, as the inverse square root diverges for small $x$. Luckily, the multiplication by $x$ after this computation takes care of the singularity and, since most bits of low-significance of $\frac 1{\sqrt x}$ will cause underflow for small $x$, we can get away with computing a shifted version of the inverse square root. This optimization reduces the number of extra bits required during the evaluation of the inverse square root.

It is worth noting that in many applications, evaluating $\arcsin(x)$ only on the interval $[0,0.5]$ may be sufficient. In such cases, the cost is much lower since this can be achieved using our parallel polynomial evaluation circuit. The Toffoli counts for this case can be found in Table~\ref{tbl:funcs}.

\subsection{Reversible implementation}
The Arcsine is implemented as a combination of polynomial evaluation and the inverse square root to extend the polynomial approximation on $[0,0.5]$ to the entire domain $[0,1]$ employing the double-argument identity above. First, the (pseudo) sign-bit of $x$ is copied out and $x$ is conditionally inverted (modulo two's-complement) to ensure $x\geq 0$. Since there are plenty of registers available, this can be achieved by conditionally initializing an extra register to $\Ket 1$ and then using a normal adder to increment $\overline x$ by one, where $\overline x$ denotes the bit- or one's-complement of $x$. Since $x\in[0,1]$, one can determine whether $x<0.5$ using just one Toffoli gate (and 4 NOT gates). The result of this comparison is stored in an ancilla qubit denoted by $\Ket a$. $z=(1-x)/2$ can be computed using an adder (run in reverse) acting on $x$ shifted by one and a new register, after having initialized it to $0.5$ using a NOT gate. Then, conditioned on $\Ket{\overline a}$ (i.e., on $a$ being 0), this result is copied into the polynomial input register $\Ket{p_{\text{in}}}$ and, conditioned on $\Ket a$, $x$ is squared into $\Ket{p_\text{in}}$. After having applied our polynomial evaluation circuit (which uncomputes intermediate results) to this input, $\Ket{p_\text{in}}$ can be uncomputed again, followed by computing the square root of $z$. Then, the result of the polynomial evaluation must be multiplied by either $\sqrt z$ or $x$, which can be achieved using $2n$ controlled swaps and one multiplier. The final transformation of the result consists of an initialization to $\pi/2$ followed by a subtraction, both conditioned on $\Ket{\overline a}$, and a copy conditioned on $\Ket a$. Finally, the initial conditional inversion of $x$ can be undone after having (conditionally) inverted the output.

\subsection{Resource estimates}
Following this procedure, the Toffoli count for this arcsine implementation on $n$-bit numbers using $m$ Newton iterations for calculating $\sqrt z$ and a degree-$d$ polynomial to approximate $\arcsin(x)$ on $[0,0.5]$ can be written as
\begin{align*}
	T_\text{arcsin}&=3T_{\text{inv}}+(2T_\text{poly}-T_\text{fma})\\
	&\phantom{={}} +2T_\text{csquare}+T_\text{mul}+T_\text{cadd}\\
	&\phantom{={}} +(2T_\text{invsqrt}+T_\text{mul})+5n+2\\
	&\phantom{={}} +T_\text{add}\\
	&=3T_\text{add}+2T_\text{poly}+3T_\text{mul}\\
	&\phantom{={}} +T_\text{cadd}+2T_\text{invsqrt}+9n+2\\
	&=d (3 n^2+n (6 p+7)-6 (p-1) p-2)\\
	&\phantom{={}}+m (n (15 n+30 p+23)-30p(p-1) -4)\\
	&\phantom{={}}+9 (n+1) p+\frac 92 n (n+1)\\
	&\phantom{={}}+6 n^2+28n-9 p^2+2
\end{align*}


where $T_\text{inv}(n)$ denotes the Toffoli count for computing the two's-complement of an $n$-bit number and $T_\text{csquare}(n,p)=T_\text{mul}(n,p)+2n$ is the number of Toffoli gates required to perform a conditional squaring operation. Furthermore, $2n$ Toffoli gates are needed to achieve the conditional $n$-bit swap operation (twice), and another $3n$ are used for (conditional) copies.

\begin{figure}[t]
	\resizebox{.95\linewidth}{!}{\input{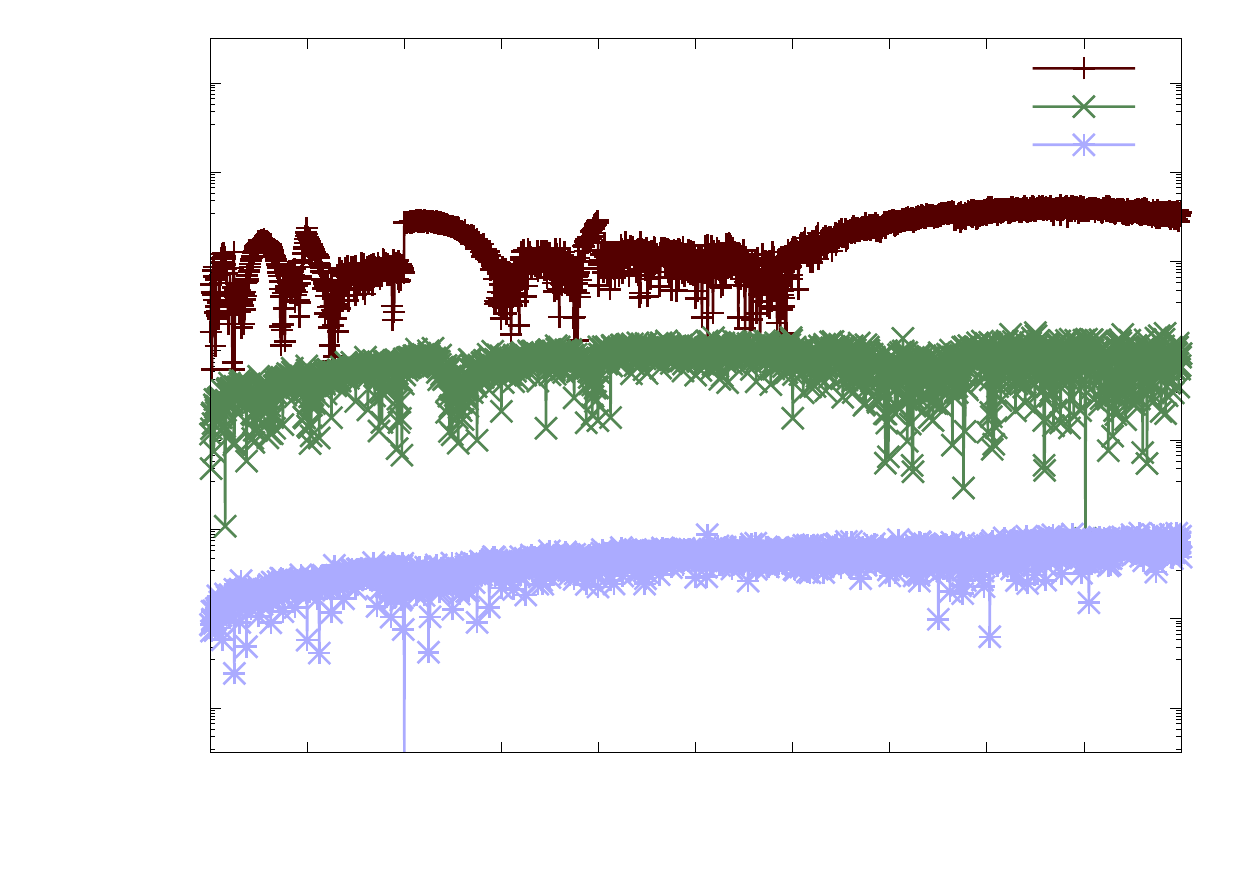}}
	\caption{Absolute error on $[0,5]$ for $N=2000$ equidistant points of our reversible implementation of the square root for $m\in\{2,3,4\}$ Newton iterations and corresponding bit sizes $n\in\{25,35,50\}$. The fixed-point position is chosen to be $p=5$.}
	\label{fig:sqrterror}
\end{figure}

\section{Results of the Reversible Simulation}

All circuits were implemented at the gate level and tested using a reversible simulator extension to LIQ$Ui\Ket{}$. The results are presented in this section.

\subsection{Piecewise polynomial approximation}
A summary of the required resource for implementing $\tanh(x)$, $\exp(-x^2)$, and $\sin(x)$ can be found in Tbl.~\ref{tbl:funcs}. For each function, one set of parameters was implemented reversibly at the level of Toffoli gates in order to verify the proposed circuits.

\subsection{(Inverse) Square root}
The convergence of our reversible fast inverse square root implementation with the number of Newton iterations can be found in Fig.~\ref{fig:invsqrterror}, where the bit sizes and point positions have been chosen such that the roundoff errors do not interfere significantly with the convergence. For all practical purposes, choosing between $3$ and $5$ Newton iterations should be sufficient. The effect of tuning the constants in the initial guess (see Eqn.~\ref{eq:initguessconstant}) can be seen when comparing Fig.~\ref{fig:invsqrterrornoopt} to Fig.~\ref{fig:invsqrterror}: The initial guess is obtained from the location of the first non-zero in the bit-representation of the input, which results in large rounding-effects for inputs close to an integer power of two. Tuning the initial guess results in almost uniform convergence, which allows to save an entire Newton iteration for a given $L_\infty$-error.

The square root converges better than the inverse square root for small values, which can be expected, since
\[
	\sqrt x = x\cdot \frac 1{\sqrt x}
\]
has a regularizing effect for small $x$. The error after $m$ Newton iterations when using $n$ bits for the fixed point representation is depicted in Fig.~\ref{fig:sqrterror}. Additionally, the initial guess could be improved by tuning the constants in Eqn.~\ref{eq:initialguess} such that the error is minimal after multiplying $x\cdot\frac 1{\sqrt x}$, instead of just optimizing for the inverse square root itself.

\subsection{Arcsine}
Our implementation of Arcsine uses both the polynomial evaluation and square root subroutines. The oscillatory behavior which can be seen in Fig.~\ref{fig:arcsinerror} is typical for minimax approximations. For $x>0.5$, the resolution is lower due to the wider range of $\frac 1{\sqrt x}$, which was accounted for by calculating a shifted version of the inverse square root. While this allows to save a few qubits (to the left of the binary point), the reduced number of qubits to the right of the binary point fail to resolve the numbers as well, which manifests itself by bit-noise for $x>0.5$ in Fig.~\ref{fig:arcsinerror}. The degrees of the minimax approximation were chosen to be $7$, $13$, and $17$ for $m=3,4,5$, respectively. Since $\arcsin(x)$ is an odd function, this amounts to evaluating a degree $3$, $6$, and $8$ polynomial in $x^2$, followed by a multiplication by $x$.

\renewcommand{\arraystretch}{1.2}
\begin{longtable*}{c|c|c|c|c|c}
		Function & $L_\infty$ error & \parbox{2cm}{Polynomial degree} & \parbox{2cm}{Number of\\subintervals} & \parbox{2cm}{Number of\\qubits} & \parbox{2cm}{Number of\\Toffoli gates}\\\hlinewd{.8pt}
		$\tanh(x)$  & & & & &\\
		 & $10^{-5}$  & & & &\\
		& & 3 & 15 & 136 & 12428\\
		& & 4 & 9 & 169 & 13768\\
		& & 5 & 7 & 201 & 15492\\
		& & 6 & 5 & 234 & 17544\\
		 & $10^{-7}$  & & & &\\
		& & 3 & 50 & 166 & 27724\\
		& & 4 & 23 & 205 & 23095\\
		& & 5 & 14 & 244 & 23570\\
		& & 6 & 10 & 284 & 26037\\
		 & $10^{-9}$  & & & &\\
		& & 3 & 162 & 192 & 77992\\
		& & 4 & 59 & 236 & 41646\\
		& & 5 & 30 & 281 & 35460\\
		& & 6 & 19 & 327 & 36578\\
		$\exp(-x^2)$  & & & & &\\
		 & $10^{-5}$  & & & &\\
		& & 3 & 11 & 132 & 10884\\
		& & 4 & 7 & 163 & 12141\\
		& & 5 & 5 & 195 & 14038\\
		& & 6 & 4 & 226 & 15863\\
		 & $10^{-7}$  & & & &\\
		& & 3 & 32 & 161 & 20504\\
		& & 4 & 15 & 199 & 19090\\
		& & 5 & 10 & 238 & 21180\\
		& & 6 & 7 & 276 & 23254\\
		 & $10^{-9}$  & & & &\\
		& & 3 & 97 & 187 & 49032\\
		& & 4 & 36 & 231 & 32305\\
		& & 5 & 19 & 275 & 30234\\
		& & 6 & 12 & 319 & 31595\\
		$\sin(x)$  & & & & &\\
		 & $10^{-5}$  & & & &\\
		& & 3 & 2 & 113 & 6188\\
		& & 4 & 2 & 141 & 7679\\
		& & 5 & 2 & 169 & 9170\\
		& & 6 & 2 & 197 & 10661\\
		 & $10^{-7}$  & & & &\\
		& & 3 & 3 & 142 & 9444\\
		& & 4 & 2 & 176 & 11480\\
		& & 5 & 2 & 211 & 13720\\
		& & 6 & 2 & 246 & 15960\\
		 & $10^{-9}$  & & & &\\
		& & 3 & 7 & 167 & 13432\\
		& & 4 & 3 & 207 & 15567\\
		& & 5 & 2 & 247 & 18322\\
		& & 6 & 2 & 288 & 21321\\
		$\exp(-x)$  & & & & &\\
		 & $10^{-5}$  & & & &\\
		& & 3 & 11 & 116 & 8106\\
		& & 4 & 6 & 143 & 8625\\
		& & 5 & 5 & 171 & 10055\\
		& & 6 & 4 & 198 & 11245\\
		 & $10^{-7}$  & & & &\\
		& & 3 & 31 & 149 & 17304\\
		& & 4 & 15 & 184 & 15690\\
		& & 5 & 9 & 220 & 16956\\
		& & 6 & 7 & 255 & 18662\\
		 & $10^{-9}$  & & & &\\
		& & 3 & 97 & 175 & 45012\\
		& & 4 & 36 & 216 & 28302\\
		& & 5 & 19 & 257 & 25721\\
		& & 6 & 12 & 298 & 26452\\
		$\arcsin(x)$  & & & & &\\
		 & $10^{-5}$  & & & &\\
		& & 3 & 2 & 105 & 4872\\
		& & 4 & 2 & 131 & 6038\\
		& & 5 & 2 & 157 & 7204\\
		& & 6 & 2 & 183 & 8370\\
		 & $10^{-7}$  & & & &\\
		& & 3 & 3 & 134 & 7784\\
		& & 4 & 2 & 166 & 9419\\
		& & 5 & 2 & 199 & 11250\\
		& & 6 & 2 & 232 & 13081\\
		 & $10^{-9}$  & & & &\\
		& & 3 & 6 & 159 & 11264\\
		& & 4 & 3 & 197 & 13138\\
		& & 5 & 3 & 236 & 15672\\
		& & 6 & 2 & 274 & 17938\\
		
		\hlinewd{1pt}
	\caption{Costs associated with the evaluation of Gaussian, hyperbolic tangent, $\sin(x)$, $\exp(-x)$ for $x\geq 0$, and $\arcsin(x)$ on $[-0.5,0.5]$ using piecewise polynomial approximation in combination with our parallel evaluation scheme. All Toffoli counts are for compute only (i.e., there is an additional factor of 2 for uncompute). For even/odd functions, the given degree corresponds to the evaluation cost, i.e., the actual polynomial being implemented has degree $2d$ or $2d+1$, respectively.}
	\label{tbl:funcs}
\end{longtable*}

\end{document}